\documentclass[a4paper,10pt]{article}
\usepackage[utf8]{inputenc}
\usepackage[T1]{fontenc}
\usepackage{tabularx}
\usepackage{lineno}
\usepackage{graphicx}
\usepackage{times}
\usepackage{hyperref}
\usepackage{subcaption}
\usepackage{listings}
\usepackage{amsmath}
\usepackage{xcolor}

\usepackage[a4paper, left=1.5cm, right=1.5cm, top=2cm, bottom=2cm]{geometry}

\definecolor{codebg}{rgb}{0.95,0.95,0.95}
\definecolor{keywords}{rgb}{0.0,0.0,0.6}
\definecolor{comments}{rgb}{0.0,0.5,0.0}
\definecolor{strings}{rgb}{0.6,0.0,0.0}

\title{Dataset resulting from the user study on comprehensibility of explainable AI algorithms}

\author{
    Szymon Bobek\textsuperscript{1,a} \and
    Paloma Korycińska\textsuperscript{2,b} \and
    Monika Krakowska\textsuperscript{3,b} \and
    Maciej Mozolewski\textsuperscript{4,a} \and
    Dorota Rak\textsuperscript{5,b} \and
    Magdalena Zych\textsuperscript{6,b} \and
    Magdalena Wójcik\textsuperscript{7,b} \and
    Grzegorz J. Nalepa\textsuperscript{8,a}
}

\date{}

\begin{document}
\maketitle

\begin{center}
    \textsuperscript{a}Jagiellonian Human-Centered AI Lab, Mark Kac Center for Complex Systems Research, Institute of Applied Computer Science, Jagiellonian University, Krakow, Poland \\
    \textsuperscript{b}Institute of Information Studies, Faculty of Management and Social Communication, Jagiellonian University, Krakow, Poland
\end{center}

\footnotetext[1]{Email: \texttt{szymon.bobek@uj.edu.pl}, ORCID: 0000-0002-6350-8405}
\footnotetext[2]{Email: \texttt{paloma.korycinska@uj.edu.pl}, ORCID: 0000-0002-4010-079X}
\footnotetext[3]{Email: \texttt{monika.krakowska@uj.edu.pl}, ORCID: 0000-0002-2724-9880}
\footnotetext[4]{Email: \texttt{m.mozolewski@doctoral.uj.edu.pl}, ORCID: 0000-0003-4227-3894}
\footnotetext[5]{Email: \texttt{dorota.rak@uj.edu.pl}, ORCID: 0000-0001-8113-9132}
\footnotetext[6]{Email: \texttt{magdalena.zych@uj.edu.pl}, ORCID: 0000-0001-9770-3674}
\footnotetext[7]{Email: \texttt{magda.wojcik@uj.edu.pl}, ORCID: 0000-0001-5059-858X}
\footnotetext[8]{Email: \texttt{grzegorz.j.nalepa@uj.edu.pl}, ORCID: 0000-0002-8182-4225}

\begin{abstract}
This paper introduces a dataset that is the result of a user study on the comprehensibility of explainable artificial intelligence (XAI) algorithms.
The study participants were recruited from 149 candidates to form three groups representing experts in the domain of mycology (DE), students with a data science and visualization background (IT) and students from social sciences and humanities (SSH). 
The main part of the dataset contains 39 transcripts of interviews during which participants were asked to complete a series of tasks and questions related to the interpretation of explanations of decisions of a machine learning model trained to distinguish between edible and inedible mushrooms.
The transcripts were complemented with additional data that includes visualizations of explanations presented to the user, results from thematic analysis, recommendations of improvements of explanations provided by the participants, and the initial survey results that allow to determine the domain knowledge of the participant and data analysis literacy.
The transcripts were manually tagged to allow for automatic matching between the text and other data related to particular fragments. 
In the advent of the area of rapid development of XAI techniques, the need for a~multidisciplinary qualitative evaluation of explainability is one of the emerging topics in the community.
Our dataset allows not only to reproduce the study we conducted, but also to open a wide range of possibilities for the analysis of the material we gathered.
\end{abstract}


\section*{Background \& Summary}


With the rapid development of black-box machine learning (ML) models, such as deep neural networks or gradient boosting trees, the need for explanations of their decisions has emerged. 
This demand has been driven by the increasing implementation of opaque models, 
in high-risk and critical areas like medicine, healthcare, industry, and law, which laid the foundation for modern research on explainable and interpretable artificial intelligence (XAI). 
Scientists' efforts in designing XAI algorithms have been further supported by political initiatives such as Defense Advanced Research Projects Agency's (DARPA) XAI challenge~\cite{darpa}, the European Union's (EU) General Data Protection Regulation (GDPR)~\cite{gdpr}, and more recently, the EU AI Act~\cite{aiact}.

The shared goal of all these initiatives is to improve the transparency of AI systems, thereby promoting their adoption in areas where trust in AI is not fully established or where the transparency of decisions is crucial for legal and safety reasons.
However, as XAI algorithms have been advanced, a new discussion has been initiated, addressing the fundamental challenge of ensuring that the explanations generated by these algorithms are comprehensible to humans.
This triggered research on the evaluation of XAI~\cite{nauta2023xaieval}, drawing attention from social sciences, which argued that much of the effort in XAI relies solely on researchers' intuition about what constitutes a \emph{good} explanation. 
They emphasized that human factors should be integral to the design and evaluation of XAI to ensure its reliability~\cite{miller2019social}.

Recognizing individual human abilities to comprehend algorithmically generated explanations is crucial, as these abilities can vary significantly based on personal information competencies.
Additionally, there is a lack of established multidisciplinary methods for measuring these capabilities, as well as datasets that facilitate reproducible evaluations or comprehensive analyses.

Our contribution addresses these gaps by presenting a dataset resulting from extensive user study on the comprehensibility of XAI algorithms across three distinct user groups with different information competencies.
The dataset contains material collected from 39 participants during the interviews conducted by the information sciences research group.
The final 39 participants were recruited from 149 candidates to form three groups that represented domain experts in the field of mycology (DE), students with a data science and visualization background (IT) and students from social sciences and humanities (SSH). 
The IT group consisted of 8 participants, the SSH group included 18 participants, and the DE group comprised 13 domain experts.
Each group was provided with a set of explanations from a machine learning model trained to predict edible and non-edible mushrooms.

The data used to train the ML model was in tabular format and is publicly available~\cite{Wagner2021MushroomDC}.
The machine learning model we used was Extreme Gradient Boosting (XGB)  and explanations for its decision were prepared including state of the art model-agnostic algorithms such as SHapley Additive exPlanations (SHAP), Local Interpretable Model-agnostic Explanations (LIME), Diverse Counterfactual Explanations (DICE) and Anchor~\cite{molnar2020interpretable}.

\begin{figure}[ht]
\centering
\includegraphics[width=\linewidth]{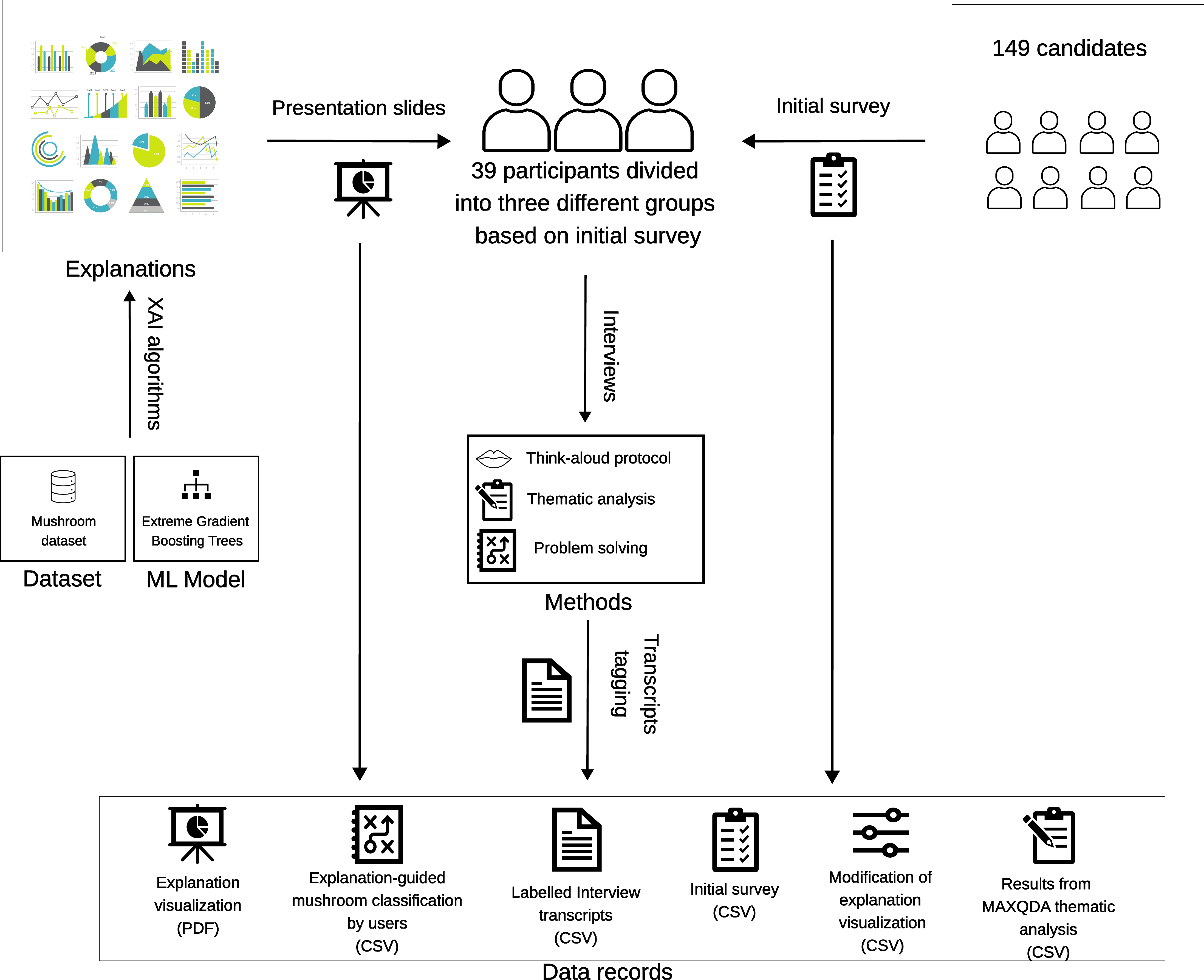}
\caption{Schema of a study, data collection and data format. The columns which contain self-explanatory headings are put in square brackets in the figure, indicating that there is a multiple columns related to some type of information, e.g. survey questions in SURVEY or feature names in PROBLEMS. }
\label{fig:workflow}
\end{figure}

During the interviews, participants were asked to interpret the presented explanations and answer related questions. 
The interviews were conducted according to the think-aloud protocol, a technique in which participants verbalize their thoughts and reactions as they interact with stimuli or perform tasks. 
This approach provides insights into participants' cognitive processes, decision-making, and understanding of the subject matter in real-time.
The transcripts gathered with this methodology capture not only the final responses of the participants but also their reasoning, uncertainties, and moments of clarity during the evaluation of XAI algorithms. The overview of the study and resulting dataset is given in Figure~\ref{fig:workflow}.

The resulting dataset combines and summarizes the whole study.
It was constructed from the self-assessment surveys obtained from the candidates, anonymized transcripts of the interviews, the results of thematic analysis of the interviews, and original explanations accompanied modifications suggested by the participants. 
The transcripts were manually tagged, allowing researchers to correlate specific text fragments with visual explanations, thematic analysis results, and user recommendations on desired process of explanation. 

This dataset addresses the gap identified in theoretical works, such as Miller's \cite{miller2019social}, which emphasize the need for empirical evaluations of XAI comprehensibility but lack real-world user studies. 
Our dataset provides a richer and more comprehensive empirical foundation by including a diverse set of user interviews, enabling a deeper understanding of how various user groups—such as domain experts, data science students, and social science students—comprehend XAI algorithms. Unlike studies such as \cite{ford2023study,nimno2024us,jalali2023us}, which focus on a specific professional group, or evaluate XAI methods within a single user group, this dataset offers a broader evaluation by examining a wider spectrum of user expertise, thus providing more comprehensive insights into the challenges XAI faces across different domains.

Our goal was to provide a dataset from the study that can serve as a resource to replicate our experiment but also for in depth analysis of the empirical material we collected. 
We envision broader applications of our dataset, such as its use in multimodal explainability methods, where interview transcripts and XAI visualizations could aid in developing and training multimodal (text-image) conversational agents for interactive XAI. We believe our dataset has the potential to significantly advance XAI research, extending far beyond the examples outlined above.
In addition, our dataset offers valuable opportunities for personalization and user modeling in XAI. The transcripts capture detailed user interactions and preferences, enabling the development of user models that reflect differences in domain knowledge, explanation needs, and cognitive strategies. These models can inform the design of adaptive XAI systems that tailor explanations to individual users, enhancing comprehension and trust. 
This makes the dataset a strong foundation for building preference-aware recommender systems and interactive interfaces that support personalized, human-centric explainability.
The dataset is complemented with the source code, allowing one to reproduce the initial ML model and explanations.


\section*{Methods}


\subsection*{Machine learning model and XAI algorithms}

The dataset used to train the classifier is a publicly available dataset from the 
University of California Irvine Machine Learning Repository~\cite{secondary_mushroom_848}. 
It contains data on 61,069 specimens from 173 mushroom species, categorized as either \textit{edible} or \textit{inedible/poisonous}. 
Specimens with unknown edibility were classified as inedible/poisonous. 
The dataset exclusively includes cap-and-stalk mushrooms with gill hymenophores and comprises both real observations and hypothetical data, the latter generated artificially from a smaller set of real-world mushroom observations.
The hypothetical and artificially generated data was created in accordance to the description of the morphological structure of the mushrooms obtained from the Mushrooms and Toadstools book ~\cite{harding_mushrooms_1999}.

Regarding class distribution, the dataset is fairly balanced, with 33,888 instances labeled inedible or poisonous and 27,181 as edible, making approximately 55.49\% of the entries \textit{inedible/poisonous}. 
This balance is crucial for accurate model training and performance evaluation, avoiding biases towards the majority class.

The ML model we prepared and presented to the study participants uses the XGB classifier. 
The motivation for XGB was its excellent performance on the dataset.
To handle categorical variables, a one-hot encoder was used and missing data was addressed through imputation. 
Additionally, numerical variables were scaled to ensure uniformity in data handling. 
This preprocessing involved imputing missing values in numeric features with the median and scaling these features, while categorical variables were imputed with a placeholder value '\_NA\_'. 
Consequently, the model operated on 82 features derived from the original 20 input variables. 
Renowned for its efficiency in classification tasks, the model exhibited an impressive accuracy rate of 99.97\%.

Study participants were informed about the model's limitations, emphasizing that while the model is highly accurate and data-driven, it does not encompass all possible factors affecting mushroom edibility. 
Therefore, it should be used as a supporting tool rather than the sole determinant in the identification process. 
Additionally, the standard practice of data partitioning was applied, dividing the dataset into a training set for model development and a test set for validation.

To cover the broadest spectrum of XAI methods without overwhelming participants, we selected the following representative types of explanations for our study:

\begin{itemize}
    \item Statistical descriptions and visualizations of data,
    \item Feature importance attribution-based explanations, such as SHAP and LIME,
    \item Rule-based explanations, such as Anchor and Local Universal Explainer (LUX)~\cite{sbk2025softlux},
    \item Counterfactual explanations, such as DICE.
\end{itemize}

The selection of explainability methods - Anchor, DICE, LIME, and SHAP - was driven by the need to cover a diverse range of well-established explainers in the field. We chose the most popular and widely recognized methods, each representing a distinct type of explanation (e.g., local, feature-based, rule-based) to ensure comprehensive coverage.

Additionally, both feature importance attribution explanations and rule-based explanations were provided in the form of local and global explanations. Our selection of methods was loosely guided by the work of Baniecki et al.~\cite{baniecki2023grammar}, which demonstrated the effectiveness of sequentially analyzing a model using a combination of multiple complementary XAI mechanisms.

\begin{figure}[ht]
\centering
\includegraphics[width=\linewidth]{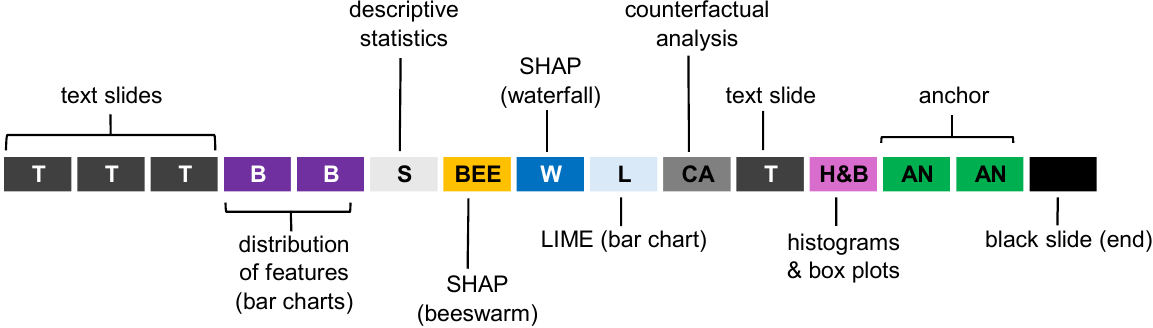}
\caption{Diagram of the slide sequence in the base presentation with types of visualization elements.}
\label{fig:ordering}
\end{figure}

The explanations were compiled in a single PDF presentation with 14 slides, where each slide contains a visualization of the explanation generated with the selected XAI algorithm. 
There was also a 15th slide that indicated the end of the study.
The type of slides and their ordering are presented in Figure~\ref{fig:ordering}.
We encourage the reader to search for the full set of slides while reading the following description of methods used to craft the presentation available on Zenodo~\cite{xaifungi}.

\subsubsection*{Statistical descriptions and visualization of data}

For each of these numerical features: 
both histograms and box plots were created using the matplotlib library from scikit-learn.
Missing data points were appropriately labeled and represented by gray bars.
The descriptive statistics for the numerical variables were presented in a straightforward tabular text output, translating all elements such as count, mean, and standard deviation into Polish. 
Categorical variables were illustrated using horizontal bar charts, with labels translated into Polish to accommodate user preferences.

\subsubsection*{Feature importance attribution-based explanations}
A "Swarm Plot" from the SHAP package offered a visual representation of the impact of individual mushroom features on the model's classification as either edible or poisonous. 
While the majority of the plot was translated into Polish, due to technical limitations, the labels "High" and "Low" on the "Feature Value" axis remained in English. 
Below the horizontal axis, a~description clarified: "SHAP Value Effect Points to the right (> 0.0) indicate a greater impact on the classification of mushrooms by the model as poisonous, to the left as edible". 
The accompanying legend is elucidated as follows: High signifies a high feature value, for instance, "cap\_diameter\_cm" (High) correlates to a large cap diameter in centimeters. 
Conversely, Low represents a low feature value, such as "stalk\_height\_cm" (Low) indicating a short stalk. 
The "0" mark delineates the boundary between features of significant and minor importance in assessing the edibility of mushrooms. 
For binary characteristics, a~high value (High), denoted by the red color, implies the presence of the feature.
In the top left corner, a succinct explanation encapsulated the essence of the "Swarm Plot": illustrating the influence of individual fungal traits on the prediction of their edibility (edible/inedible or poisonous). 

A SHAP "Waterfall" chart was utilized to interpret the model's feature attributions for prediction. 
The custom title indicated the analysis of features that influence the prediction of the specified class, with the true class of the specimen also noted. 
This visualization delineated the contribution of each feature to the model's prediction, with positive values indicating increased probability of the mushroom being classified as poisonous, and negative values the opposite. 
The chart was accompanied by a~legend where: $E[f(X)]$ signifies the baseline value or the mean model prediction, the color bands reflect the influence of each feature on the prediction of toxicity for the individual mushroom, and $f(x)$ is the final value equating to the model's prediction for that specific mushroom. 
An explanatory note stated the waterfall chart's function as displaying the impact of individual features on the toxicity prediction of a given mushroom. The term $E[f(X)]$ denoted the model's average prediction across all observations, while $f(x)$ represents the prediction for the specific instance in question. 

A tailored LIME feature importance chart was created, averaging out the attributes that consistently influenced the model's predictions. 
Features with an average impact near zero were omitted for the sake of clarity. 
The chart highlighted features in red that generally contributed to a mushroom being classified as non-edible or poisonous, and in blue for those indicating edibility. 
LIME provides granular feature importance, such as "stalk\_height\_cm > 7.74" or specific colors for "stalk\_color", which were reformulated into more readable conditions like "stalk = {color}" for interpretability.
The slide's was titled: "LIME Chart: Feature Values Affecting the Probability of Predicting Edibility". 
The horizontal axis denotes the likelihood shift towards edibility or toxicity due to each feature. 
Since the LIME package does not include this specific chart type, it was derived by calculating the average importance across successive random batches of observations until reaching a convergence threshold of 0.01. 
This process ensured that only features with a significant average contribution were displayed, filtering out those whose influence varied too much across observations.

\subsubsection*{Rule-based explanations}
The study employed the ANCHOR package, a method of explainability for AI models that generates an "anchor", or a set of conditions that collectively influence the model's classification of an instance with high precision. 
An example observation matching the anchor criteria was displayed on the left side of the slide, representing the output from Anchor. 
The central part of the slide described the AI's prediction process: for instances meeting all listed anchor conditions, the AI would predict the class as EDIBLE with 97.2\% certainty. 
The anchor, composed of feature conjunctions that determine classification, is detailed in the upper right. 
Below it, the extent of influence these conditions had on the certainty of the classification was quantified.
Since interactive features were not employed, the slide showed a non-interactive list of features. 
Replicating the checklist with fewer marked features offered insights into how different subsets contributed to the prediction, without providing full anchor conditions. 
Two slides were presented, one for the edible class and one for the inedible/poisonous class. 

\subsubsection*{Counterfactual explanations}
Counterfactual analysis was depicted through a textual output table on the left, listing features along with their original values and the altered values required to reverse the AI's prediction. 
On the right, the explanatory text illustrated the interpretation of this analysis, detailing how AI-filling of missing data in the description of a specific mushroom influences the prediction of its edibility. 
The specimen in question, originally found in nature as non-edible/poisonous, was correctly assigned by the AI to the predicted non-edible/poisonous class. 
The counterfactuals demonstrated the minimal data alterations needed for the mushroom to be predicted as edible.
The legend clarified: 
\textit{Original Value}: Data obtained for the specific mushroom from the original dataset,
\textit{Modified Value}: Data altered or supplemented by the AI model.
The dice\_explainer from the respective package was employed for calculating counterfactuals. 
The presentation excluded cases where the 'missing data' value was changed, as such instances were considered non-informative for interpretation purposes.

\subsection*{Collection of empirical material}
The collection of empirical material comprised two main phases: an online survey and individual qualitative interviews conducted using the think-aloud protocol (TAP) guidelines~\cite{someren1994tap}. 
The survey involved non-intrusive questioning, while the TAP interviews required audio recording with explicit consent obtained from willing participants. 
Participation in both the survey and interviews was voluntary, and all collected data was anonymized.
Before starting the survey, participants were asked to confirm that their participation was voluntary and that they acknowledged the data collected during the experiment would be shared for scientific purposes, in accordance to GDPR regulations.
Participants were also asked for their consent to make all empirical material - including interview transcripts and other data collected during the study - available for publication after anonymization.
We opted for general professional vocabulary over industry-specific jargon to ensure clear communication without adding new content. 

In accordance with the guidelines approved by the Council of the Faculty of Management and Social Communication for the ethics committee's procedures, each research team is responsible for independently assessing whether their study requires submission for committee review.
Based on this self-assessment, our research did not meet the criteria for ethics committee review for the following reasons: a) We explicitly stated in both the consent forms and verbally (with recordings available) that the focus of the study is not on evaluating individuals, but rather on the material presented to participants (i.e., types of explanations generated by an artificial intelligence method); and b) The research poses no psychological harm or discomfort, as confirmed in our documentation.

Initially, the research aimed to recruit participants willing to undergo interviews, assessing their knowledge, qualifications, experience, and competencies in the specific domain relevant to the training of AI models and eXplainable AI (XAI) methods, focusing on mushroom classification in our case. 
The survey results categorized participants into three aforementioned groups: domain experts (DE), those familiar with data analysis (IT), and those with no data analysis expertise (SSH).

Subsequently, these groups were assigned tasks and interviewed using advanced tools from social science fields such as TAP and MAXimum Qualitative Data Analysis (MAXQDA) for analysis. 
The interview part of the study included collection of qualitative material for further analysis with the TAP procedure.
The MAXQDA methodology was chosen due to its widespread use in qualitative and mixed-methods research, particularly in user studies, for systematically analyzing and interpreting textual and multimedia data. It also enables the integration of qualitative insights with quantitative data, supporting the development of more user-centered and interactive explainability methods.

In the TAP study, we employed thematic analysis to systematically explore user perspectives, identify recurring themes, and uncover meaningful insights that quantitative methods alone might overlook. This approach is especially well-suited for capturing the depth and complexity of user experiences, which is crucial for understanding how users interact with AI explanations.

A total of 26 students (IT and SSH) and 13 experts took part in TAP.  
The IT group had 8 respondents from computer science, applied computer science, computer game science and electronic information processing. 
The SSH group therefore consisted of 18 respondents representing the following fields of study: psychology, practical psychology, applied psychology, information management, contemporary culture lab, journalism and social communication, and cognitive science. 
The SSH group was more extensive, as we were particularly interested in exploring the interaction experiences with XAI techniques of respondents not professionally involved in data analysis and visualisation in the context of artificial intelligence and computer science. 

Regarding the tasks performed by the respondents during TAP, we utilized a list of three main tasks.
In the Task 1~participants were asked to review a list of questions related to each explanation and answer as many as they could. 
There were 17 questions in total, each corresponding to a specific explanation type.
Additionally, there were three problem-solving questions where participants were asked to classify mushrooms based on the knowledge they gained from the explanations.
In the second task (Task 2), participants were asked to arrange the explanations in the order they found most useful, removing any unnecessary explanations and possibly adding general ideas on what explanations should be included and their preferred order.
In the third task (Task 3), participants were asked a general question: "What improvements can be made to the prepared materials to make them more useful to you?"

Task 1 was the longest one, as it was the most elaborate due to a set of detailed questions mainly aimed at evaluating individual XAI techniques. 
Task 2 and 3 mainly served to elicit recommendations for improving the prepared XAI explanations. 
Additionally, Task 2 was modelled on the card sorting technique known in UX, which facilitates the design of the information architecture of information systems~\cite{najafgholinejad2022ux}.  

We ran the research using TAP from January to March 2024. 
In the case of students (IT and SSH groups), the first session took place on 17.01.2024 and the last session on 19.02.2024. 
All sessions with students were conducted stationary in the building of the Faculty of Management and Social Communication at the Jagiellonian University. 
Each time, a single study was conducted in a room (the researcher's office room or the laboratory room) with only the researcher, the respondent and eventually one observer. 
We recorded the material obtained from the sessions with the students in audio using Microsoft Teams, with an automatic transcription option in Polish. 
In order to avoid possible data loss related to the reliability of MS Teams, we also recorded all sessions with students using a backup tool: an audio recorder on the phone or a voice recorder. 
We further processed the automatically generated transcriptions manually in order to properly capture the terminology and identify the voices of the researcher and respondent.
Finally, we labeled different fragments of the text to distinguish between different tasks and questions that the text refer to.    

In the case of the experts (DE), all sessions were carried out remotely via MS Teams, and audio and video were recorded. 
Again, automatic transcription into Polish was used, which was then manually corrected for terminological clarity. 
In the case of remote TAP sessions, however, there was no need to manually identify voices, as these were automatically signed by Microsoft Teams. 
TAP research with experts lasted from 01.02.2024 to 04.03.2024.

The TAP study was followed by a thematic analysis which is a qualitative data analysis approach that emphasizes identifying and describing prominent themes within the data. 
It also typically involves comparing themes across different conditions or populations and exploring the relationships between these themes~\cite{puryear2016inside,josephsen2017qualitative,rojas2020qualitative}.
Thematic analysis comprises of six consecutive stages.
The first stage involved an exploratory phase for data reconnaissance and familiarization. 
The second phase consisted of the independent, inductive generation of codes by two researchers.
Next, themes were created and verified, during which the initial and subsequent codebooks were developed by  deductive referencing codes to the primary research goals of our research which was focused on evaluation of comprehensibility of XAI.
This stage includes refinement of original codes and themes once the analysis progress.
Finally, the analysis concluded with the compilation of results, highlighting the co-occurrences of themes among different observations.~\cite{braun2019tap}.

Between January 17, 2024, and March 4, 2024, a total of 66 items were collected for coding, including respectively: 13 transcripts of research video sessions with domain experts, as well as 27 transcripts of study conducted with students and 26 notes prepared by the think-aloud protocol facilitators after the study's conclusion. 
MAXQDA 2024 Analytics Pro software (the UJ licence is valid until 5 January 2025) was used for coding the materials and their initial analysis and visualisation, as was Tableau Software for visualising the results. 
The dataset presented in our work contains both the codebook designed during the analysis as well as the final result of the analysis in a form of codes' occurrences among different participants.

\section*{Data Records}


The complete dataset was deposited in the Zenodo repository~\cite{xaifungi}.
The general structure of the dataset is described in Table~\ref{tab:file_description}. 
The files that contain in their names [RR]\_[SS]\_[NN] contain the individual results obtained from particular participant. 
The meaning of the prefix is as follows:

\begin{itemize}
\item RR - initials of the researcher conducting the interview,
\item SS - group that the participant belongs to (DE for domain expert, SSH for social sciences and humanities students, or IT for computer science students),
\item NN - the participant number representing the order in which the participants were interviewed by a given researcher.
\end{itemize}

The dataset contains a mix of Polish and English content. 
The components created during the post-processing phase of the study were prepared in English, while the core components such as the transcripts and slides with explanations were left in their original Polish form to preserve the integrity of the results obtained. Additionally, given the rapid advances in language processing and translation, we believe that the dataset retains more value in its original language, allowing for more accurate translations with increasingly powerful tools when necessary.

\begin{table}[ht]
\centering
\begin{tabularx}{\textwidth}{|l|X|l|}
\hline
\textbf{File} & \textbf{Description} & \textbf{Language} \\
\hline
\texttt{SURVEY.csv} & Results from a survey filled out by 143 participants, with 39 selected to form the final participant group. & Polish \\
\hline
\texttt{SURVEY\_EN.csv} & Translation of the SURVEY content to English. & English \\
\hline
\texttt{CODEBOOK.csv} & Codebook used in thematic analysis and MAXQDA coding. & English \\
\hline
\texttt{QUESTIONS.csv} & List of questions asked to participants during interviews. & Polish and English \\
\hline
\texttt{SLIDES.csv} & List of slides used in the study, including their interpretation and reference to MAXQDA themes and VISUAL\_MODIFICATIONS tables. & English \\
\hline
\texttt{MAXQDA\_SUMMARY.csv} & Summary of thematic analysis with codes from CODEBOOK used for each participant. & English \\
\hline
\texttt{PROBLEMS.csv} & List of problems presented to participants during interviews, corresponding to instances in the dataset for classification. & Polish \\
\hline
\texttt{PROBLEMS\_EN.csv} & Translation of the PROBLEMS content to English & English \\
\hline
\texttt{PROBLEMS\_RESPONSES.csv} & Responses to the problems listed in PROBLEMS.csv for each participant. & Polish and English \\
\hline
\texttt{VISUALIZATION\_MODIFICATIONS.csv} & Information on modifications made to the order of slides, removal of slides, and suggested additional explanations by participants. & English \\
\hline
\texttt{ORIGINAL\_VISUALIZATIONS.pdf} & PDF file containing original visualizations presented to participants. & Polish \\
\hline
\texttt{ORIGINAL\_VISUALIZATIONS\_EN.pdf} & PDF file containing original visualizations presented to participants translated into English. & English \\
\hline
\texttt{VISUALIZATION\_MODIFICATIONS.zip} & ZIP file containing modified slides from ORIGINAL\_VISUALIZATIONS.pdf, each named with participant ID (e.g., [RR]\_[SS]\_[NN].pdf). & Polish \\
\hline
\texttt{TRANSCRIPTS.zip} & ZIP archive containing anonymized interview transcripts for each participant, named after their ID (e.g., [RR]\_[SS]\_[NN].csv). Transcripts include text tagged with slide, question, and problem numbers. & Polish \\
\hline
\end{tabularx}
\caption{Description of files related to the dataset}
\label{tab:file_description}
\end{table}

\begin{figure}[ht]
\centering
\includegraphics[width=\linewidth]{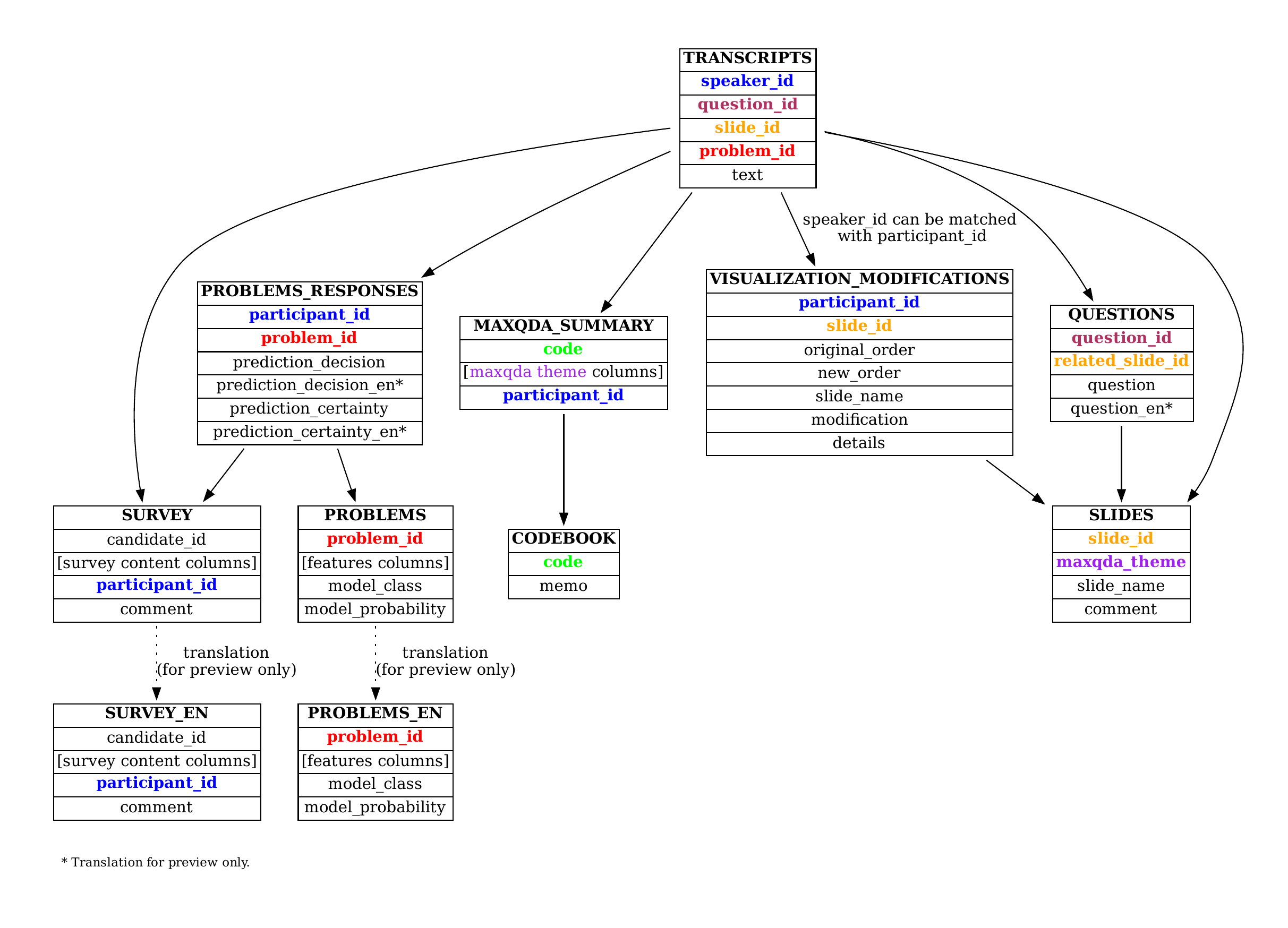}
\caption{Scheme of data records. Arrows represent intended way of joining tables, while colors denote keys in different tables that can be used to join tables in other ways. Tables containing translations should be used for preview only to better understand the content. Column name between square brackets mean that it actually represents a set of columns which names are self-explanatory and there were not included in the figure for the sake of readability.}
\label{fig:schema}
\end{figure}

The central component of the dataset is the \texttt{TRANSCRIPT} file, which includes manually tagged transcripts of interviews with participants. It can be linked to other tables using unique identifiers, thereby enriching the interviews with additional selected data. The diagram illustrating the relationships between dataset components is shown in Figure~\ref{fig:schema}.

\subsection*{Initial survey results}

The initial survey dataset contains self-assessment questions that all participant candidates were required to answer. 
The survey comprised 18 questions, 12 of which pertained directly to self-assessing knowledge and skills in mushrooming and data visualisation, as well as acquiring information about their origin and experience. The survey included a question about the field of study to ensure appropriate representation of students pursuing humanities, social sciences, and computer science. Five questions in the questionnaire directly related to giving consent for data use and participation in further research, as well as confirming familiarity with the GDPR clause. The questionnaire aimed to gather information and assess the knowledge, qualifications, experience, and competences of the participants. It consisted of six open-ended questions and six closed questions. The latter included an 'other' option, allowing participants to provide their own answer. 
These questions enable the grouping of participants based on their domain knowledge and data analysis literacy, which is crucial for selecting the final set of participants and categorizing them into DE, IT, and SSH groups.
The description of a content of this set of data is given in Table~\ref{tab:survey_columns}.
The survey table contains 154 records, including 4 candidates used in the pilot study who were not included in the final dataset. The goal of the pilot study was to identify potential issues, refine research methodologies, and ensure the feasibility of the study design; therefore, the participants in this stage were not considered valid subjects for the final dataset.
The missing values in this table indicate that the candidate did not provide an answer. Candidates who were not selected for interviews do not have a participant\_id assigned.

\begin{table}[ht]
\centering
\begin{tabularx}{\textwidth}{|l|X|}
\hline
\textbf{Column name} & \textbf{Column description} \\
\hline
candidate\_id & Unique ID of the participant of a survey. Note that this ID is not used further in other files. Instead, \texttt{participant\_id} should be used to join records from the CSV files. \\
\hline
[set of columns corresponding to survey content] & These columns are self-explanatory; each column title represents a question asked to participants, and the content provides their answers. Additionally, these columns may include metadata related to the survey, such as time spent filling out the survey and dates of completion. \\
\hline
participant\_id & Unique identifier of a participant who took part in the interviews. This column should be used to join records from other tables. Candidates with missing ID were not selected for the interviews.\\
\hline
comment & The additional comment related to the entry.\\
\hline
\end{tabularx}
\caption{Description of columns in the survey dataset}
\label{tab:survey_columns}
\end{table}

\subsection*{Labelled interview transcripts}

The core component of the dataset consists of interview transcripts with manually labeled text fragments. 
Each fragment was assigned to a slide with a specific explanation visualization (see Table~\ref{tab:slide_columns}), the question the participant is trying to answer (see Table~\ref{tab:question_columns}), and the problem the participant is attempting to solve (see Table~\ref{tab:mushroom_classification_columns}). 
Additionally, special tags were used to divide the transcripts into stages corresponding to the three tasks performed by the participants: explanation analysis, problem-solving, and visualization modifications.
In the dataset, each transcript is represented by a separate file, named with a unique participant ID key that is also present in other tables. 
The schema description of the transcripts is provided in Table~\ref{tab:dataset_description}.

\begin{table}[ht]
\centering
\begin{tabularx}{\textwidth}{|l|X|}
\hline
\textbf{Column name} & \textbf{Column description} \\
\hline
speaker\_id & ID of a person who authored the text in the following column. This distinguishes between investigators and participants. \\
\hline
slide\_id & ID of a slide that matches the ID in SLIDES.csv to which the following text is related. The row where the slide ID appears marks the starting point where this slide is displayed to the participant, or where the particular stage of the interview begins or ends.  \\
\hline
question\_id & ID of a question that can be matched with QUESTIONS.csv. It represents where the question was asked or where the participant started giving an answer. Not all questions from QUESTIONS can be matched with transcripts due to some participants not answering certain questions or vague tagging of question answers. \\
\hline
problem\_id & ID of a problem that participants were asked to solve, which can be matched with PROBLEMS.csv. The appearance of this ID in a row indicates that from this point onward, the participant attempted to solve the problem. \\ 
\hline
text & Anonymized transcript of the participant's words obtained from MS Teams. \\
\hline
\end{tabularx}
\caption{Description of columns in the transcripts files}
\label{tab:dataset_description}
\end{table}

The key column to identify the stage of the interview is \texttt{slide\_id} column.
Besides the ID of a currently analyzed slide, there are also three other special IDs that mark different parts of the interview.
The interpretation of the special IDs is as follows: \texttt{\_\_S00\_\_} indicates the beginning of the core part of the interview with slide analysis, before that the description of the study is introduced to the participant; \texttt{\_\_S99\_\_} represent the beginning of the section where the participants analyze visualization order, the slides ids are not assigned in this section, due to dynamic slide switching by participants; \texttt{\_\_S15\_\_} represents the end of the slide analysis section; \texttt{\_\_S88\_\_} represents the beginning of the problem solving section.

\begin{table}[ht]
\centering
\begin{tabularx}{\textwidth}{|l|X|}
\hline
\textbf{Column name} & \textbf{Column description} \\
\hline
slide\_id & Unique ID of a slide that allows joining with other tables using slide ID, such as \texttt{TRANSCRIPTS}. \\
\hline
maxqda\_theme & The name of the explanation type used in thematic analysis, present in columns of \texttt{MAXQDA\_SUMMARY.csv}. \\
\hline
slide\_name & Slide name used in \texttt{VISUALIZATION\_MODIFICATIONS.csv}. \\
\hline
comment & Explanation or content description of the slide. \\
\hline
\end{tabularx}
\caption{Description of columns related to slides in the dataset}
\label{tab:slide_columns}
\end{table}

\begin{table}[ht]
\centering
\begin{tabularx}{\textwidth}{|l|X|}
\hline
\textbf{Column name} & \textbf{Column description} \\
\hline
question\_id & Unique ID of each question, allowing matching of the actual question text with transcripts. \\
\hline
related\_slide\_id & The slide ID that the question was originally assigned to. \\
\hline
question & The actual text of the question. \\
\hline
question\_en & Translation of the question to English\\
\hline
\end{tabularx}
\caption{Description of columns related to questions}
\label{tab:question_columns}
\end{table}

\subsection*{Explanation-guided mushroom classification by users}
The participants were asked to solve three problems related to mushroom classification as edible and non-edible based on the explanations and other material they were provided.
All of the participants were asked about the same three instances that contain two edible and one non-edible mushroom with varying ML model classification probability assigned to each instance.
The table containing this instances along with ML model predictions is described in detail in Table~\ref{tab:mushroom_classification_columns}, while the responses of the participants were stored in a table of a schema defined by Table~\ref{tab:problem_response_columns}.
The participants were also asked to indicate their level of certainty of prediction that can later be confronted with ML model classification probability.
Missing values in this table indicates that the participant did not provide an answer. 
The DE group of participants was not solving this tasks, hence the table is missing participant IDs related to domain experts in mycology.

\begin{table}[ht]
\centering
\begin{tabularx}{\textwidth}{|l|X|}
\hline
\textbf{Column name} & \textbf{Column description} \\
\hline
problem\_id & Unique ID of the participant that can be used to join with other tables. \\
\hline
[set of features for a given mushroom] & These columns represent the features of a mushroom used in a particular problem. \\
\hline
Features values & Values of the features of the mushroom to be classified. \\
\hline
model\_class & Class returned by the machine learning model. \\
\hline
model\_probability & Probability assigned by the machine learning model to the prediction. \\
\hline
\end{tabularx}
\caption{Description of columns related to mushroom classification problems}
\label{tab:mushroom_classification_columns}
\end{table}

\begin{table}[ht]
\centering
\begin{tabularx}{\textwidth}{|l|X|}
\hline
\textbf{Column name} & \textbf{Column description} \\
\hline
problem\_id & Unique ID of the problem that can be used to join the responses with the \texttt{PROBLEMS} table and other tables. \\
\hline
participant\_id & Unique ID representing a participant, used as a key to join with other tables. \\
\hline
prediction\_decision & The class assigned by the participant to the given problem. \\
\hline
prediction\_decision\_en & Translation to English of the class assigned by the participant to the given problem. \\
\hline
prediction\_certainty & The certainty of the participant's decision. \\
\hline
prediction\_certainty\_en & Translation to English of the certainty of the participant's decision. \\
\hline
\end{tabularx}
\caption{Description of columns related to problem responses}
\label{tab:problem_response_columns}
\end{table}

\subsection*{Modification of explanation visualization}
One of the tasks assigned to participants was to suggest modifications to the provided explanations to improve overall user experience and enhance their usability. 
Participants could reorder the explanations, remove selected ones, or add custom slides with additional explanations or information they deemed useful. 
The schema of the table containing records from this part of the study is shown in Table~\ref{tab:slide_modification_columns}.
The DE group of participants was not solving this tasks neither. 
Missing values in the slide\_id or original\_order column indicate that the slide was not part of the original presentation and was suggested for addition by the participant.  
Missing values in the new\_order column indicate that the slide was recommended for removal.

\begin{table}[ht]
\centering
\begin{tabularx}{\textwidth}{|l|X|}
        \hline
        \textbf{Column name} & \textbf{Column description} \\
        \hline
        participant\_id & Unique ID of the problem that can be used to join the table with other tables. \\
        \hline
        slide\_id & ID of a slide used in other tables.  \\
        \hline
        original\_order & The original order of a given slide in the presentation. In case of custom slides, they were added to the presentation by the participant, this field is empty. \\
        \hline
        new\_order & The order (place in a presentation) of a slide assigned by the participant. \\
        \hline
        slide\_name & Symbolic name of the slide. \\
        \hline
        modification & The type of modification suggested by the participant, where 0 means no modification, 1 removal, and 2 addition of custom slide. \\
        \hline
        details & Details of a custom slide added to the presentation. In the case of the other slides, this field is empty. \\
        \hline
\end{tabularx}
\caption{Description of columns related to slide modifications}
\label{tab:slide_modification_columns}
\end{table}

\subsection*{Results from thematic analysis with MAXQDA}
The transcripts were additionally analysed with the MAXQDA thematic analysis toolkit.
The codes used for the thematic analysis are stored in a file defined by a schema presented in Table~\ref{tab:thematic_analysis_columns}.
The final codebook contained both overarching and subordinate codes related to: evaluation criteria (such as credibility, accuracy, comprehensibility, importance), methods of analysis (including thoroughness, variability of decisions or evaluations, difficulties), data and features (such as feature significance, understanding of data gaps, lack of appropriate data, understanding of annotations), knowledge and experience in analysing human behaviour, knowledge and experience in visualizations and analysis, knowledge and experience in fungi (including codes regarding names and species of fungi mentioned by participants), affects (e.g. interest, reluctance, uncertainty, surprise), names of slides used in the study (e.g. waterfall, anchor, counterfactual analysis), aesthetics (including subordinate codes regarding font size, colour scheme, layout), recommendations (e.g. addition of supplementary materials, graphics), as well as separate codes regarding trust, barriers (such as concerns, limitations), expectations, personification, metaphorization, and associations.

The final result of the analysis with occurrences of particular codes within participants' transcripts are stored in a format presented in Table~\ref{tab:code_occurrences_columns}.
Due to notable subject-matter differences in research material obtained from DE group, comparatively to the text corpus provided by IT and SSH groups, the backbone code tree was extended by adding several subordinate codes, without altering the structure of the overarching ones.
Therefore, the results for IT and SSH groups do not include all the codes present in the codebook.

\begin{table}[ht]
\centering
\begin{tabularx}{\textwidth}{|l|X|}
\hline
\textbf{Column name} & \textbf{Column description} \\
\hline
code & Name of the code that can be matched with \texttt{CODEBOOK.csv}. \\
\hline
[list of columns corresponding to particular type of explanation] & These columns represent different types of explanations, such as LIME, descriptive statistics, etc. \\
\hline
occurrences & Number of occurrences of the code in the respective type of explanation. \\
\hline
participant\_id & ID of the participant for which the summary was prepared. \\
\hline
\end{tabularx}
\caption{Description of columns related to code occurrences in explanations}
\label{tab:code_occurrences_columns}
\end{table}

\begin{table}[ht]
\centering
\begin{tabular}{|l|p{0.7\linewidth}|}
\hline
\textbf{Column name} & \textbf{Column description} \\
\hline
code & The name of a code used in thematic analysis that includes hierarchical levels separated by >. For example, \texttt{Aesthetics > layout} represents a code on the second level with \texttt{Aesthetics} as its parent. \\
\hline
memo & The meaning or description associated with a given code in thematic analysis. \\
\hline
\end{tabular}
\caption{Description of columns in thematic analysis dataset}
\label{tab:thematic_analysis_columns}
\end{table}

\section*{Technical Validation}

\subsection*{Classification model training and validation}
The creation of a ML model was followed by the standard practices in data science including numerical data scaling, categorical variables encoding and missing values imputation.
The dataset was balanced, so not oversampling nor under-sampling techniques were applied that could alter the original data distribution.
The model training was performed on train set which comprises 70\% of all samples randomly drawn for the full dataset.
The hyper-parameter tuning was performed with grid search and five-fold cross validation and tested on the test-set not used for any training task before, to mitigate the overfitting risks.
With this approach, we achieved a model performance of 99.97\% accuracy, which is consistent with results reported in other studies using this dataset~\cite{secondary_mushroom_848}
For the full transparency and reproducibility of this procedure, we provide a source code of the model training and data preprocessing (see: \url{https://gitlab.geist.re/pro/xai-fungi}).

\subsection*{Study design and data acquisition }
To ensure data quality during the empirical material acquisition stage, we addressed potential issues on two levels. 
The first level focused on creating a representative sample of participants for the study, while the second level involved ensuring the high quality of the material collected during the study.

To verify that the sample of participants recruited for the study was representative and diverse enough to distinguish significantly different groups with respect to their background knowledge in the domains of macro-fungi and data science, we undertook several steps.
Firstly, we designed a comprehensive survey, developed by social sciences researchers with expertise in information sciences, to capture user profiles from an information capabilities perspective. 
This survey allowed us to accurately assess the background knowledge of participants in the domains of macro-fungi and data science. 
To reach a diverse group of students, we advertised the research across three different universities in Poland, thereby broadening our participant base. 
Additionally, domain experts were recruited from the mycological society, specifically among individuals who are certified fungi identifiers accredited by the competent Regional Office of Sanitary and Epidemiological Vigilance (SANEPID) in Poland. 
This selection ensured that expertise among the domain experts was not solely based on self-assessment, which could be unreliable, but rather confirmed by their official certification from a recognized legal authority in Poland.

The total number of enrolled participants (N = 39) concords with the evidence-based recommendations on sample size in phenomenological studies~\cite{boddy2016sample,vasileiou2018sample,hennik2022sample}. 
The data saturation, evidenced by the repetitiveness of codes and codes’ co-occurrence patterns, has been achieved across all three subsets of respondents under study. 
Furthermore, the sample size is in alignment with, and in fact exceeds, the size of the most recent published studies employing the Think-Aloud Protocol to examine XAI interpretations from a variety of end-user categories~\cite{anjara2023tap,li2024tapmining,rodgers2024procol}.

To ensure the high quality of empirical material obtained during interviews, we identified and addressed the limitations and drawbacks inherent in the TAP procedure.
Firstly, due to the need to verbalize, it is a demanding technique for the respondents, at the same time as the situation itself may seem unnatural. 
We aimed to minimise this potential discomfort with the following actions: (1) we showed the respondents an audio recording of an exemplary verbalisation so that they could more easily immerse themselves in the procedure (the recording was made on a website unrelated to the subject of our study, so as not to suggest to the respondents a direction for interpreting the materials), (2) the current question/task was kept on a sheet of paper in the respondent's sight at all times, so that the respondents could more easily realise where they were in the research and which question they were answering, (3) in the case of longer periods of silence, the interviewer reminded the respondent of the verbalisation in as neutral way as possible (e.g. what are you currently looking at?), (4) we provided mineral water for the respondents to drink, (5) we incorporated into the procedure the possibility of a break during the session at the request of the respondent. 

Additionally, the TAP interviews with students have been conducted in comfort-ensuring conditions. 
The rooms devoted to one-to-one sessions have been silent and isolated from external disturbances. The privacy of the respondents has been preserved. 
Everyone has been offered a bottle of mineral water and the possibility of requesting a break at any moment. 
The sessions were audio-recorded on two devices to minimize the risk of data loss. The interviews with domain experts took the form of video meetings on the MS Teams platform and were video-recorded with automatic transcription in Polish. In both cases (students \& experts) the automatic transcripts were reviewed and corrected manually by the researcher who carried out the given session. 
The coding was accomplished by three coders who worked in coordination, agreeing on successive rectifications and extensions of the coding tree. 
Due to the specific content revealed in domain experts’ verbal output, the coding tree for experts was added with several new items.


\subsection*{Data post-processing}
The post-processing of the results, involving the transformation of all gathered material into structured CSV format as presented in previous sections, underwent subsequent verification steps. 
Initially, the technical integrity of the material was assessed by two researchers to ensure no data records were missing from the original sources. 
Following this, the transcripts and their manual labeling were reviewed by three independent researchers to confirm the accuracy and integrity of this stage.
Finally, we performed technical analysis of the dataset, ensuring that all of the tables contain valid IDs and can be merged according to schema depicted in Figure~\ref{fig:schema}.
The part of the dataset that were translated into English, were translated with a help of ChatGPT, and verified by us manually.

\section*{Usage Notes}



In the online repository (see \url{https://gitlab.geist.re/pro/xai-fungi}), we present how to read sample records from a dataset, how to link the records with other tables and trigger basic analysis with \emph{scikit-learn} library~\cite{scikitlearn}.
The examples assume that all of the files are in the \texttt{zenodo} folder and the transcripts were unzipped into \texttt{zenodo/transripts} folder.







\section*{Code availability}


The source code used for preprocessing of the dataset used to train the classifier, the actual training procedure of the classifier, and the generation of explanations used in the study is available at: \url{https://gitlab.geist.re/pro/xai-fungi}.


\section*{Acknowledgements} 
The paper is funded from the XPM project funded by the National Science Centre, Poland under the CHIST-ERA programme grant agreement Np. 857925 (NCN UMO-2020/02/Y/ST6/00070).
The research has been supported by a grant from the Priority Research Area (DigiWorld) under the Strategic Programme Excellence Initiative at Jagiellonian University.

\section*{Author contributions statement}

All authors contributed equally to the work;

\section*{Competing interests} 

Apart from the funding, the authors declare no competing interests. 

\bibliographystyle{ieeetr}
\bibliography{xaifungi}

\end{document}